\documentclass[journal,10pt]{IEEEtran}
\usepackage{cite}
\usepackage{amsmath,amssymb,amsfonts}
\usepackage{algorithmic}
\usepackage{graphicx}
\usepackage{textcomp}
\usepackage{caption}
\usepackage{subcaption}
\usepackage{xcolor}
\usepackage{mathtools}
\usepackage{orcidlink}
\usepackage{hyperref}
\usepackage{pgfplots}
\usepackage{verbatim}
\usepackage{float}

\usepackage{graphicx}
\usepackage{subcaption}
\usepackage{circuitikz}
\usepackage{amsmath}

\begin{document}

\title{Hardware Limitations and Optimization Approach in 1-Bit RIS Design at 28 GHz}

\author{
Hossein Rezaei\orcidlink{0009-0007-8696-3025}, Mehmet Emin Arslan\orcidlink{0000-0001-6913-0389}, George Yammine\orcidlink{0000-0003-2690-5855}, Niels Neumann\orcidlink{0000-0002-9112-5890} and
Norman Franchi\orcidlink{0000-0002-2777-4722}
\thanks{H. Rezaei and G. Yammine are with Fraunhofer Institute for Integrated Circuits IIS, 90411 Nürnberg, Germany.}
\thanks{M. E. Arslan and N. Neumann are with Institute for Electrical Information Technology, Clausthal University of Technology, 38678 Clausthal-Zellerfeld, Germany.}
\thanks{N. Franchi is with Institute for Smart Electronics and Systems, Friedrich-Alexander-Universität
Erlangen-Nürnberg, 91058 Erlangen, Germany.}}



\maketitle

\begin{abstract}
Reconfigurable intelligent surfaces (RIS) have emerged as a transformative technology for electromagnetic (EM) wave manipulation, offering unprecedented control over wave reflections compared to traditional metallic reflectors. By utilizing an array of tunable elements, RIS can steer and shape electromagnetic waves to enhance signal quality in wireless communication and radar systems. However, practical implementations face significant challenges due to hardware limitations and phase quantization errors.
In this work, a 1-bit RIS prototype operating at 28 GHz is developed to experimentally evaluate the impact of hardware constraints on RIS performance. Unlike conventional studies that model RIS as an ideal phase-shift matrix, this study accounts for physical parameters that influence the actual reflection pattern. In particular, the presence of specular reflection due to hardware limitations is investigated. Additionally, the effects of phase quantization errors, which stem from the discrete nature of RIS elements, are analyzed, and a genetic algorithm (GA)-based optimization is introduced to mitigate these errors. The proposed optimization strategy effectively reduces gain degradation at the desired angle caused by 1-bit quantization, enhancing the overall performance of RIS. The effectiveness of the approach is validated through measurements, underscoring the importance of advanced phase control techniques in improving the functionality of RIS.
\end{abstract}

\begin{IEEEkeywords}
Electromagnetic reflection, genetic algorithms,
hardware, optimization,
reconfigurable intelligent surfaces    
\end{IEEEkeywords}

\section{Introduction}
\label{sec: Intro}

The primary objective of developing reconfigurable intelligent surfaces (RIS) is to enhance the control over the direction of reflected electromagnetic (EM) waves, which surpasses the capabilities of traditional metallic reflectors governed by Snell's law\cite{jackson1998classical, Kaina2014, Zhu2013}. Traditional reflectors direct an incident wave either at an angle equal to the angle of incidence, as seen with mirrors, or at a fixed angle in the case of non-uniform surfaces, offering limited flexibility in controlling the direction of the reflected wave.

An innovative RIS design employs individually controllable unit cells that can be strategically positioned and configured to perform phase adjustments. This configuration allows for the manipulation of the wavefront, thereby enabling the reflection of waves in specific, predetermined directions and angles\cite{wu}. Such a level of control is pivotal for various applications in wireless communications and radar technology, where directing signals precisely is crucial\cite{holograph}. When properly configured, RISs can enable secure wireless backhaul \cite{ris_sdn}, improve channel rank \cite{channel}, and enhance localization accuracy \cite{ris_positioning}, among other advantages.

However, the effectiveness of RIS technology is not only determined by its design, but also significantly influenced by the underlying hardware. The choice and quality of the materials and components that can be used are critical to the performance of the system. Innovations in hardware technology make it possible to achieve the precise control required for RIS to function optimally. 

The tuning of resonant frequencies in RIS phase shifters can be achieved by various tuning mechanisms. The most commonly used methods are circuit tuning, geometric tuning, and material tuning  methods \cite{turpin2014reconfigurable}. The circuit tuning method uses components such as PIN diodes, varactor diodes, and RF switches for the desired phase shifts. The geometric tuning method changes the physical properties of conductive elements such as micro-electro-mechanical systems (MEMS), while the material tuning method uses substances such as liquid crystals, which involves adjusting the permittivity, permeability and conductivity of these materials.
In hardware design, PIN diodes, varactor diodes, MEMS, RF switches, relay switches, and liquid crystals are commonly used \cite{rana2023review}. RF switches and varactor diodes offer tunability for sub~6~GHz  applications, but face problems such as high losses at higher frequencies. Therefore, they are mainly used in sub~6~GHz applications. MEMS, liquid crystal materials and relay switches are mainly in the mmWave range. MEMS are compact and efficient but complex to manufacture; liquid crystals allow continuous phase shifting but have slow response times; relay switches provide high isolation but have slow switching speeds. 
PIN diodes are preferred for sub~6~GHz and mmWave applications due to their low power consumption, fast switching times, high linearity, minimal losses at high frequencies, and robustness that overcomes their limited phase resolution \cite{li2020novel, gros2021reconfigurable}. Therefore, PIN diodes are most widely used for their reliability and efficiency at high frequency advantages.

Despite advancements in hardware design, the phase shift introduced by each reflective element within the RIS is usually limited to specific values due to hardware constraints. Therefore, optimizing these phase shifts is critical to fully realizing the advantages of this technology and ensuring the system operates with maximum efficiency and precision.

While discrete phase shift optimization for RIS has received growing attention, many studies approach the problem by first relaxing the discrete constraint into a continuous one. The relaxed problem is solved using continuous optimization methods, and the result is then quantized to the nearest discrete phase value \cite{discrete_ris}. Within this framework, several commonly used optimization techniques for continuous RIS phase shift design include semidefinite relaxation \cite{9685349,9235486}, the minorize-maximization algorithm \cite{15}, penalty methods \cite{13,18}, manifold optimization \cite{19,20}, the alternating direction method of multipliers \cite{16,17}, and approaches that treat phase shifts as optimization variables rather than the complex gains they produce \cite{21,22}. Additionally, artificial intelligence based techniques, such as unsupervised learning \cite{unsupervised}, supervised learning \cite{supervised}, and reinforcement learning \cite{RL}, have recently emerged as promising solutions.

Recently, increased focus has been directed toward the discrete phase-shift model, driven by the argument that hardware constraints limit reflecting elements to a finite number of reflection levels. Compared to its continuous counterpart, the resulting resource allocation problem becomes even more complex, as it involves both continuous and discrete variables. Most studies address this challenge by adopting quantization. When the number of discrete phase shifts is sufficiently large, the performance loss from quantization remains minimal, which explains its widespread adoption. However, if the number of available phase shifts is small, the quantization method inevitably leads to performance degradation. In the field of discrete RIS optimization problem, several optimization techniques were considered. Exhaustive search \cite{23}, while exact, becomes impractical due to its exponential time complexity with increasing RIS size. Methods based on continuous relaxation followed by projection \cite{28}, offer efficient solutions in the relaxed domain but suffer from performance degradation when projected back to the discrete phase set. Branch-and-bound methods \cite{24,25} guarantee optimality under certain pruning strategies, but similarly become computationally intensive for moderate to large-scale problems. Particle swarm optimization \cite{26}, though effective in continuous settings, exhibits instability and premature convergence in discrete binary scenarios without careful tuning. Clearly, solving the discrete phase shift design problem is still in its early stages, and developing a low-complexity method that maintains strong performance remains a key challenge.

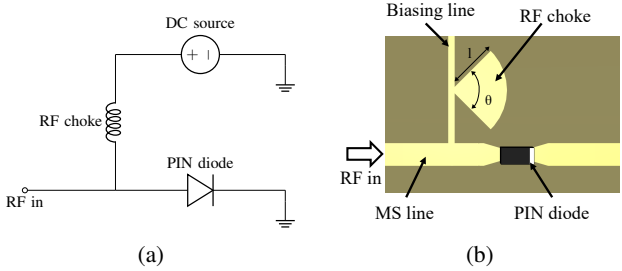
\begin{figure}[t]
  \centering
  \begin{subfigure}[b]{0.22\textwidth}
    \centering
    \resizebox{1\textwidth}{!}{%
    \begin{circuitikz}
    \tikzstyle{every node}=[font=\LARGE]
    \draw (5.75,6.75) to[short, -o] (3.75,6.75) node[below] { \normalsize RF in};
    \draw (5.75,6.75) to[L, l={ \normalsize RF choke} ] (5.75,9.75);
    \draw (5.75,9.75) to[american voltage source,l={ \normalsize DC source}] (9.5,9.75);
    \draw (9.5,9.75) to (9.5,9.5) node[ground]{};
    \draw (5.75,6.75) to[D, l={ \normalsize PIN diode}] (9.5,6.75);
    \draw (9.5,6.75) to (9.5,6.5) node[ground]{};
    \end{circuitikz}
    }
    \caption{}
    \label{fig: 1a}
  \end{subfigure}
  \quad
  \begin{subfigure}[b]{0.21\textwidth}
    \centering
    \includegraphics[width=\textwidth]{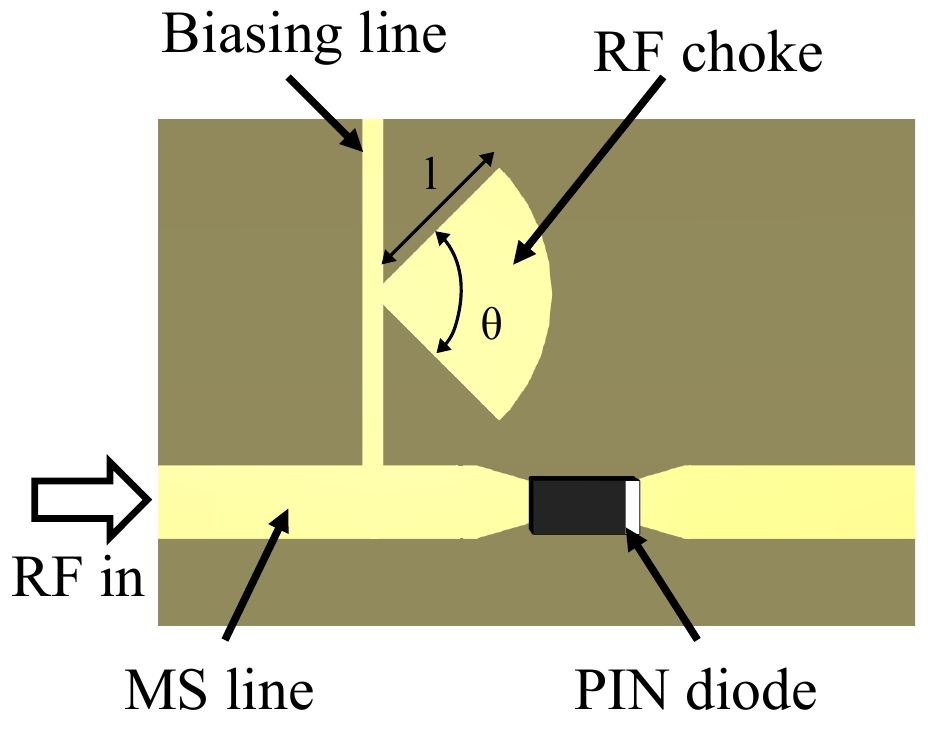}
    \caption{}
    \label{fig: 1b}
  \end{subfigure}
  \caption{Biasing network. a) Schematic of network, b) Layout of network components.}
  \label{fig:fig1}
\end{figure}

In this work, we develop a 1-bit RIS hardware prototype operating at 28 GHz. Using this setup, we account for the physical parameters that influence the actual reflection pattern of the RIS, distinguishing our study from previous works that primarily model RIS as a phase-shift matrix in mathematical formulations. Unlike conventional studies that optimize this matrix solely based on theoretical constraints, our approach incorporates practical hardware limitations, particularly those affecting mmWave operation, which have been largely overlooked in existing literature.

Prior works on discrete optimization of RIS, e.g., \cite{23,28,24,25,26}, typically treat RIS as an integrated component of the communication system, optimizing its phase distribution based on constraints such as beamforming vectors, artificial noise, and transmit power. In contrast, we consider RIS as an independent entity within the system and focus on maximizing its reflection power based on the incident angle of incoming signals and the desired reflection direction. Our optimization strategy prioritizes RIS-specific parameters that shape its reflection pattern, rather than relying on external constraints like channel state information, transceiver beamforming, or power allocation.
This perspective provides a new direction for RIS design, emphasizing practical implementation challenges and expanding the understanding of RIS behavior in real-world scenarios.

The paper is structured as follows.  Section \ref{sec: unit cell} details the unit cell design and phase shift control. Section \ref{sec: Natural ref} addresses hardware limitations and examines its impact on the reflection properties of the surface. Section \ref{sec: phase optimization} investigates the impact of phase distribution and quantization error introduced by 1-bit phase shifters, and proposes a GA-based optimization approach. Section \ref{sec:Measurement} presents the measurement setup and verifies the effectiveness of the proposed approach. Section \ref{sec: conclusion} concludes the paper.

\section{Unit Cell Design}
\label{sec: unit cell}

\begin{figure}[t]
    \centering
    \begin{subfigure}[h]{0.2\textwidth}
        \centering
        \includegraphics[scale=0.3]{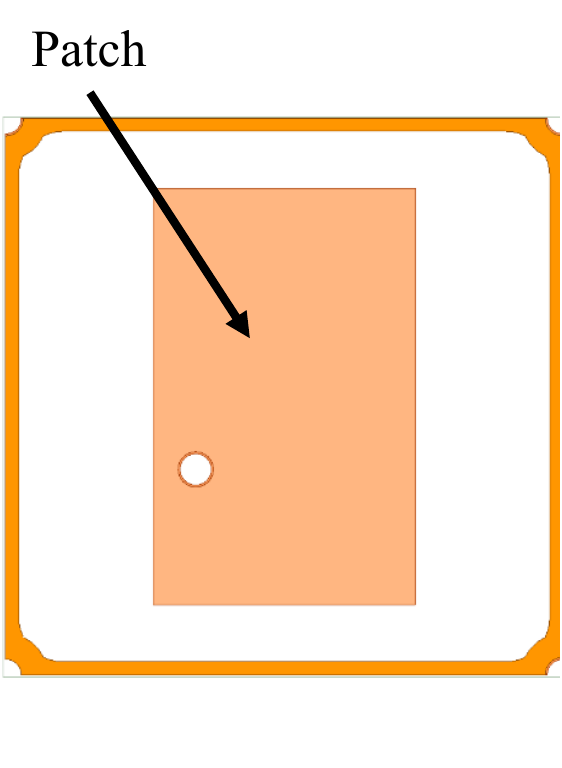}
        \caption{}
        \label{fig:unit_cell_design}
    \end{subfigure}
    \hfill 
    \begin{subfigure}[h]{0.2\textwidth}
        \centering
        \includegraphics[scale=0.3]{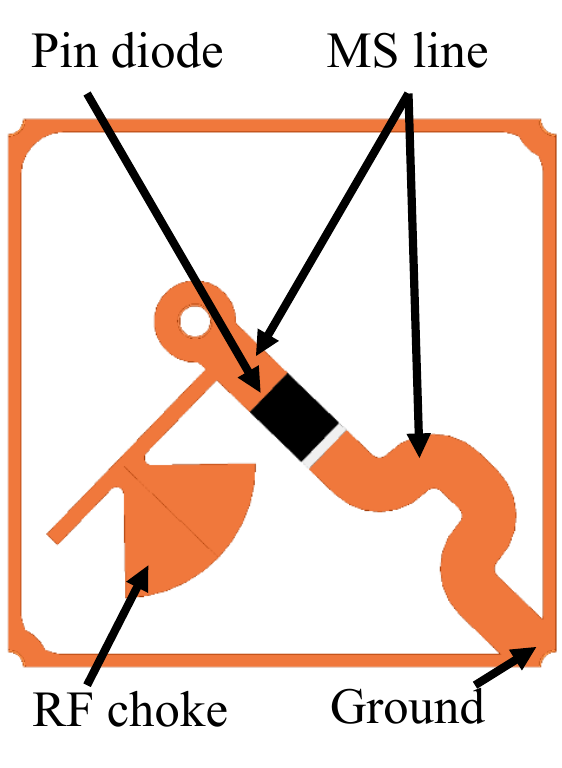}
        \caption{}
        \label{fig:unit_cell_design2}
    \end{subfigure}
    \caption{Unit cell design for a 28 GHz RIS with patch antenna and biasing network. a) top view, b) bottom view.}
    \label{fig:unit_cells}
\end{figure}

Commonly, RISs consist of hundreds to thousands unit cells \cite{wu,ris2}. Each unit cell is composed of two main components: an antenna and a biasing network. The antenna component, with its inherent gain and directivity, captures incoming EM waves. These waves are then processed by the biasing network, which introduces a controllable phase shift to tune the reflective properties of the unit cell. This phase shift functions as an RF switch, allowing the trajectory of the wave to be dynamically reconfigured. By coordinating the operation of all unit cells within the array, RISs can achieve beamforming gain, significantly enhancing the capability to direct and reflect waves as desired. Through this interaction between the antenna and the biasing network, RISs dynamically manipulate the propagation of EM waves.

\begin{figure}[t]
    \centering
    \begin{subfigure}[t]{0.2\textwidth}
        \centering
        \includegraphics[scale=0.3]{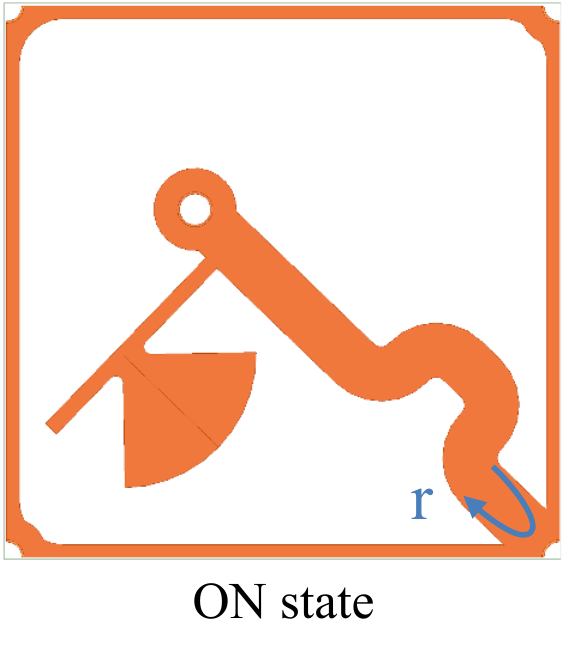}
        \caption{}
    \end{subfigure}
    \hfill 
    \begin{subfigure}[t]{0.2\textwidth}
        \centering
        \includegraphics[scale=0.3]{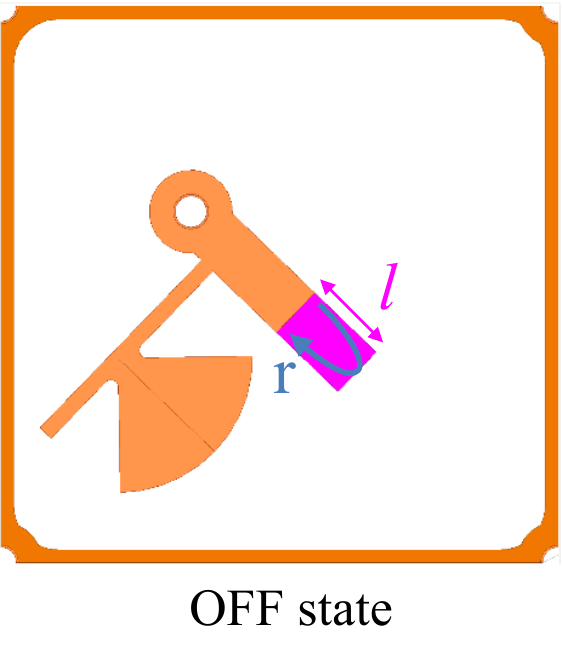}
        \caption{}
        \label{off state}
    \end{subfigure}
    \caption{Equivalent unit cell models. (a) ON state, (b) OFF state.}
    \label{fig: equivalent unit cell}
\end{figure}

The study by \cite{arslan20242x2, ris_patent} designed a tile suitable for the antenna component of a RIS unit cell, consisting of four patch antennas. The authors used a three-layer patch antenna geometry: the patch is on the top layer, the middle layer serves as a ground plane, and the bottom layer contains a feed point that can be replaced by a bias network in a unit cell design.  

To achieve the phase shift of a unit cell, a biasing network with a PIN diode is used, as shown in Fig.~\ref{fig: 1a}. The diode changes the impedance characteristics of an EM wave transmitted through the microstrip (MS) line. The characteristics of the diode can be adjusted by applying a bias from a DC source. An RF choke is used to prevent high frequencies from entering the DC path while maintaining very low resistance for DC currents.

\par The design of the biasing network is critical to optimize the performance of the unit cell. Fig.~\ref{fig: 1b} shows the layout of the biasing network components. The track width of the MS line is determined using standard methods, which involve iteratively adjusting for impedance matching based on substrate properties \cite{hammerstad1980accurate}. The geometry and placement of the RF choke is determined using a parametric study in ANSYS HFSS. The MS line behind the PIN diode is connected to the ground to enable the reflection of the transmitting EM wave. A Macom MADP-000907-14020 PIN diode along with the components shown in Fig.~\ref{fig: 1b}, have been characterized on a de-embedding board to determine the diode's impedance characteristics. The PIN diode features variable impedance depending on its bias condition. In the OFF (unbiased) state, the diode has a nearly infinite impedance, reflecting incoming EM waves. In the ON state (biased), the impedance drops to nearly zero, allowing the transmission of EM waves that are then reflected at ground. This transition between the ON and OFF states provides the required phase shift in the biasing network.

Fig.~\ref{fig:unit_cells} shows a model of a unit cell designed to generate a phase shift at an operating frequency of 28~GHz. This model integrates both the antenna and the biasing components. The antenna element, specifically the patch, is positioned on the top surface of the unit cell as shown in Fig.~\ref{fig:unit_cell_design}, while the biasing components are located on the bottom surface, as shown in Fig.~ \ref{fig:unit_cell_design2}.

\begin{figure}[t]
    \centering
    \includegraphics[scale=0.6]{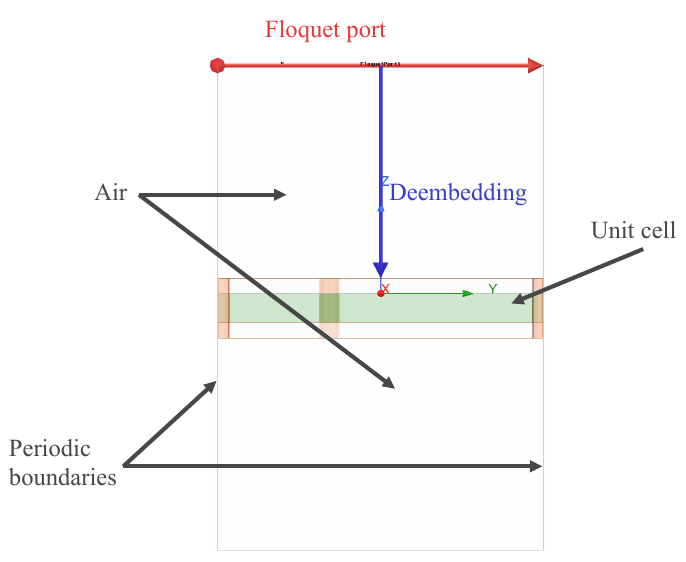}
    \caption{Setup for the FEM simulation of a unit cell in ANSYS-HFSS.}
    \label{fig: fem setup}
\end{figure}

Due to the complexity involved in characterizing the effects of the PIN diode, the presented biasing network shown in Fig.~\ref{fig:unit_cell_design2} can be simplified using equivalent models that take into account the impedance characteristics of the characterized PIN diode. Simplifying the diode helps to analyze and design the performance of the unit cell more effectively. Two equivalent models have been developed to represent the ON and OFF states of the unit cell, as shown in Fig.~\ref{fig: equivalent unit cell}. In the ON state, the PIN diode is represented by the same MS line because it does not change the impedance of the incident and reflected EM waves. The end of the MS line is connected to the ground to facilitate the transmission and reflection of the waves. Conversely, in the OFF state, the end of the MS line is left open, simulating the behavior of the unbiased PIN diode, which reflects incoming waves due to its high impedance.

To achieve different phase changes, the length of the MS line in the OFF state, denoted by \( l \), is adjusted, as shown in Fig.~\ref{off state}. During this adjustment, the design of the MS line in the ON state remains unchanged.

To accurately tune the unit cells and determine the desired phase shift between the ON and OFF states depending on the PIN diode position, simulations are performed using ANSYS HFSS which is essential for simulating EM interactions within complex structures. In these simulations, a Floquet port is used to excite the unit cell and measure its reflection coefficient. To model the behavior of an infinite array, periodic boundary conditions are applied to the lateral surfaces. Fig.~\ref{fig: fem setup} shows the unit cell configuration within the Finite Element Method (FEM) simulation. The Floquet port is de-embedded in the direction indicated by the blue arrow, and periodic boundary conditions are applied to the lateral surfaces. To mitigate external effects, the air height in the $z$ and $-z$ directions is set to a minimum of $\lambda/4$.

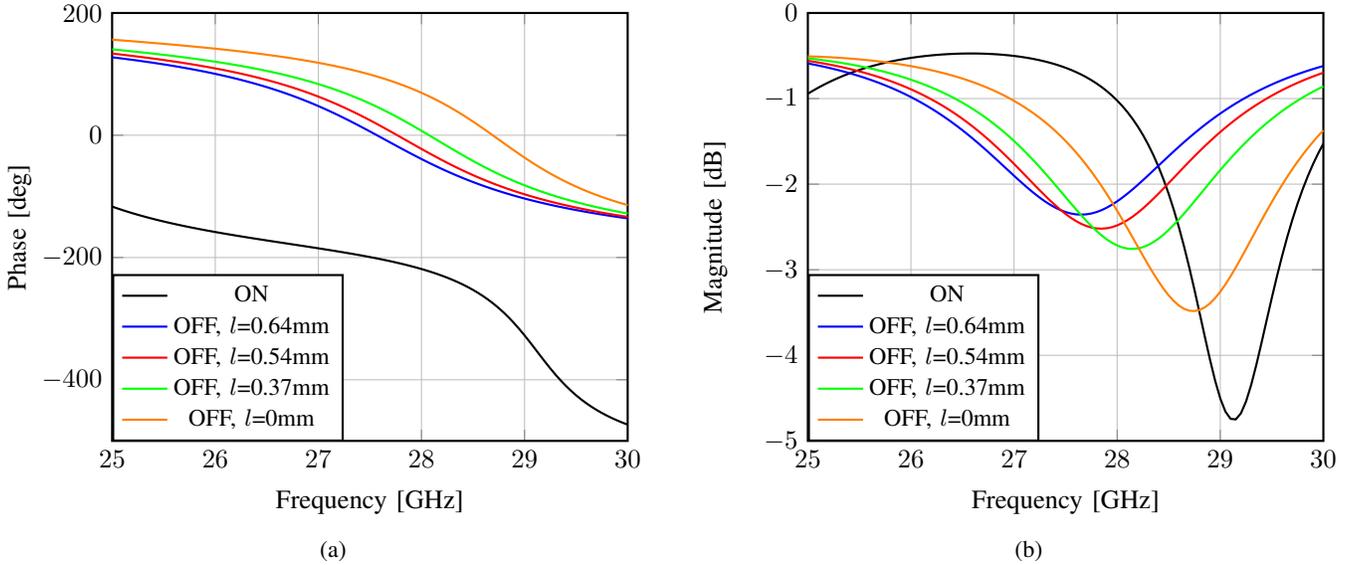
\begin{figure*}[t]
\centering
\begin{subfigure}[b]{0.49\textwidth}
    \begin{tikzpicture}
        \begin{axis}[
            xlabel={Frequency [GHz]},
            ylabel={Phase [deg]},
            legend style={
                at={(0,0.39)}, 
                anchor=north west,
                font=\small, 
            },
            xmin=25, xmax=30,
            ymin=-500, ymax=200,
            grid=major,
            thick
            ]

            \addplot[black, thick] table [x index=0, y index=1, col sep=comma] {Data/5a.csv};
        \addlegendentry{ON}
        
        \addplot[blue, thick] table [x index=0, y index=2, col sep=comma] {Data/5a.csv};
        \addlegendentry{OFF, \( l \)=0.64mm}
        
        \addplot[red, thick] table [x index=0, y index=3, col sep=comma] {Data/5a.csv};
        \addlegendentry{OFF, \( l \)=0.54mm}
        
        \addplot[green, thick] table [x index=0, y index=4, col sep=comma] {Data/5a.csv};
        \addlegendentry{OFF, \( l \)=0.37mm}

        \addplot[orange, thick] table [x index=0, y index=5, col sep=comma] {Data/5a.csv};
        \addlegendentry{OFF, \( l \)=0mm}

        \end{axis}
    \end{tikzpicture}
    \caption{}
    \label{fig:second}
\end{subfigure}
\hfill
\begin{subfigure}[b]{0.49\textwidth}
    \begin{tikzpicture}
        \begin{axis}[
            xlabel={Frequency [GHz]},
            ylabel={Magnitude [dB]},
            legend style={
                at={(0,0.39)}, 
                anchor=north west,
                font=\small, 
            },
            xmin=25, xmax=30,
            ymin=-5, ymax=0,
            grid=major,
            thick
            ]

            \addplot[black, thick] table [x index=0, y index=1, col sep=comma] {Data/5b.csv};
        \addlegendentry{ON}
        
        \addplot[blue, thick] table [x index=0, y index=2, col sep=comma] {Data/5b.csv};
        \addlegendentry{OFF, \( l \)=0.64mm}
        
        \addplot[red, thick] table [x index=0, y index=3, col sep=comma] {Data/5b.csv};
        \addlegendentry{OFF, \( l \)=0.54mm}
        
        \addplot[green, thick] table [x index=0, y index=4, col sep=comma] {Data/5b.csv};
        \addlegendentry{OFF, \( l \)=0.37mm}

        \addplot[orange, thick] table [x index=0, y index=5, col sep=comma] {Data/5b.csv};
        \addlegendentry{OFF, \( l \)=0mm}

        \end{axis}
    \end{tikzpicture}
    \caption{}
    \label{fig:first}
\end{subfigure}

\caption{Phase and reflection magnitude of unit cells in the ON and OFF states for varying MS lengths.  (a) Reflection phase. (b) Reflection magnitude.}
\label{fig:S parameter and phase shift}
\end{figure*}

As illustrated in Fig.~\ref{fig:S parameter and phase shift}, adjusting the length \( l \) of the MS line allows control over the phase difference between the ON and OFF states of the unit cell. For example, at an operating frequency of 28~GHz, lengths of \( l \) = 0.64~mm, \( l \) = 0.54~mm, \( l \) = 0.37~mm, and \( l \) =~0 mm correspond to phase differences of 180°, 164°, 134° and 72°, respectively. The magnitude of the reflected waves at this frequency ranges from approximately -1 to -2.7~dB.


\section{Hardware Limitations}
\label{sec: Natural ref}

Hardware limitations considerably influence the performance and efficacy of RIS. In practical scenarios, it is challenging to apply an independently desired phase shift to each of these cells due to technological restraints. Most RIS designs use 1-bit phase shifters for each unit cell \cite{1bit, 1bit_2, 1bit_3}, which restricts each cell to just two possible states, significantly limiting the configurability of the surface. A common RIS design uses PIN diodes to control each unit cell in a binary manner which simplifies hardware and lowers costs.

In a 1-bit RIS tuned by PIN diode, each unit cell is designed to ideally provide a phase difference of \(180^\circ\) between its ON and OFF state. However, achieving the desired phase shift can be challenging due to tolerances in hardware fabrication, which can significantly affect performance, especially as operating frequencies increase. 

For instance, a unit cell operating at \(28\, \text{GHz}\) typically exhibits physical dimensions of approximately \(5.35\,\text{mm}\) in both width and height, corresponding to the conventional $\lambda/2$ spacing. The compact size of these unit cells complicates the optimization of their performance while also posing challenges in maintaining the mechanical stability of the components in a limited area for higher frequencies. Consequently, only a 1-bit phase shifter using PIN diodes can be practically implemented at this frequency range. This restriction impacts the quality of the reflected wave from the RIS.

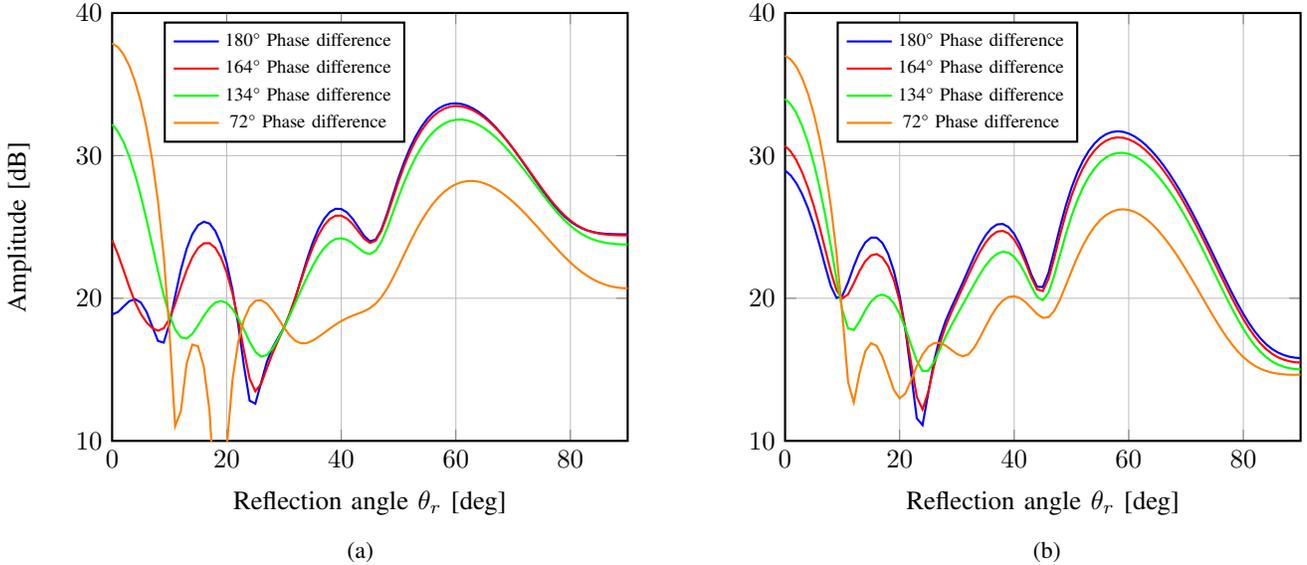
\begin{figure*}[ht]
\centering
\begin{subfigure}[b]{0.53\textwidth}
    \begin{tikzpicture}
        \begin{axis}[
            xlabel={\text{Reflection angle} $\theta_r$ [deg]},
            ylabel={Amplitude [dB]},
            legend style={
                at={(0.1,0.98)}, 
                anchor=north west,
                font=\scriptsize, 
            },
            xmin=0, xmax=90,
            ymin=10, ymax=40,
            grid=major,
            thick
            ]

        \addplot[blue, thick] table [x index=0, y index=4, col sep=comma] {Data/output.csv};
        \addlegendentry{180° Phase difference}

        \addplot[red, thick] table [x index=0, y index=3, col sep=comma] {Data/output.csv};
        \addlegendentry{164° Phase difference}

        \addplot[green, thick] table [x index=0, y index=2, col sep=comma] {Data/output.csv};
        \addlegendentry{134° Phase difference}
        
        \addplot[orange, thick] table [x index=0, y index=1, col sep=comma] {Data/output.csv};
        \addlegendentry{72° Phase difference}
        
        \end{axis}
    \end{tikzpicture}
    \caption{}
    \label{fig:first}
\end{subfigure}
\hfill
\begin{subfigure}[b]{0.45\textwidth}
    \begin{tikzpicture}
        \begin{axis}[
            xlabel={\text{Reflection angle} $\theta_r$ [deg]},
            legend style={
                at={(0.1,0.98)}, 
                anchor=north west,
                font=\scriptsize, 
            },
            xmin=0, xmax=90,
            ymin=10, ymax=40,
            grid=major,
            thick
            ]

            \addplot[blue, thick] table [x index=0, y index=1, col sep=comma] {Data/phase_difference.csv};
        \addlegendentry{180° Phase difference}
        
        \addplot[red, thick] table [x index=0, y index=2, col sep=comma] {Data/phase_difference.csv};
        \addlegendentry{164° Phase difference}
        
        \addplot[green, thick] table [x index=0, y index=3, col sep=comma] {Data/phase_difference.csv};
        \addlegendentry{134° Phase difference}
        
        \addplot[orange, thick] table [x index=0, y index=4, col sep=comma] {Data/phase_difference.csv};
        \addlegendentry{72° Phase difference}

        \end{axis}
    \end{tikzpicture}
    \caption{}
    \label{fig:second}
\end{subfigure}
\hfill

\caption{Effect of phase difference between the ON and OFF states of unit cells on amplitude of reflections. (a) Analytical simulation results. (b) FEM simulation results.}
\label{fig:phase diff. Analy. and FEM}
\end{figure*}

The effectiveness of a RIS depends on the quality of the reflections, and the overall performance of a RIS can be degraded by unintended reflections. In \cite{specular_reflection}, the impact of specular reflection from the surface behind the RIS (e.g., walls) has been considered. However, in this study, unintended reflections caused by the RIS components themselves have been investigated.

These undesigned reflections can be attributed to two primary sources: the back layer and the front layer design. Each of these parts plays a crucial role in manipulating the incident signals, and their imperfections or design limitations can lead to unwanted signal reflections, interfering with the intended functionality of the RIS. In this part, we explore how these factors impact signal behavior and examines their effects on the overall performance of the RIS. 

\subsubsection{Back Layer Effect}
The back layers of a RIS can unintentionally reflect signals in undesigned directions, including specular angles, due to various physical and environmental factors. While the front layer of a RIS is engineered with phase-shifting elements to control wavefronts precisely, the back layer lacks these controls and may act as a passive reflector. This can result from the presence of passive components, such as connectors or electronic circuits, which scatter EM waves in practically unpredictable ways. Additionally, if the back layer includes metallic elements, such as ground planes or circuits, it may behave like a conventional mirror, reflecting signals as specular reflections. Both effects may superimpose. However, unlike the controlled reflections on the front, these reflections are not configurable and can propagate in unintended directions. The absence of proper absorptive materials or insufficient shielding can exacerbate this issue, leading to unwanted signal propagation. Such reflections can cause interference with other communication systems and degrade the overall performance of the RIS by introducing multipath interference, highlighting the need for careful design and shielding of the back layer in RIS implementations.

\subsubsection{Front Layer Design}

In the design of 1-bit phase-shifting unit cells, the primary concern is not the absolute value of the phase shift exhibited by each element, but rather the phase difference between two distinct operational states: the ON and OFF states. Ideally, unit cells should be engineered to achieve a phase difference as close to $180^{\circ}$ as possible. Deviations can lead to a reduction in the desired design performance. Specifically, a divergence from the ideal phase difference results in reduced gain in the intended reflection and increased reflection in the specular direction.

To demonstrate the effects of non-ideal phase differences, we consider a RIS operating at 28~GHz comprising an $11\times11$ array of unit cells, where each unit cell has physical dimensions of $\lambda/2 \times \lambda/2$. Additionally, the center-to-center distances between unit cells are set to $\lambda/2$. In this illustrative example, the incident wave is directed normally towards the surface, and the surface itself is designed exemplarily to reflect this wave at a 60° angle.

\par Following the initial discussion on phase differences, we present the results from both analytical simulations and FEM simulations in Fig.~\ref{fig:phase diff. Analy. and FEM}. Analytical simulations typically focus solely on the front layer of the RIS, neglecting the influence of the underlying layers. In contrast, full-wave FEM simulations capture the complete multilayer structure, enabling a more accurate assessment of the electromagnetic behavior by considering the interactions between all constituent layers. These simulations were conducted for various phase difference values, allowing us to observe the effects of non-ideal phase differences. As illustrated in Fig.~\ref{fig:phase diff. Analy. and FEM}, the study considered four distinct values for the phase difference of the unit cells' states, specifically 180°, 164°, 134°, and 72°. The amplitude of the reflected wave is evaluated as a generalized form of the radar cross section of an object \cite{balanis2012}. As anticipated, by increasing the phase differences deterioration from 180° to 72° correlates with a reduction in the reflection gain in the designed direction at 60°. Concurrently, there is an increase in specular reflection observed at 0°. This trend highlights the sensitivity of the reflection characteristics to changes in the phase difference of the unit cells. As previously mentioned, any deviation from the ideal 180° phase difference leads to reduced gain in the designed reflection and an increase in specular reflections.
\par As seen in Fig.~\ref{fig:second}, even with a 180° phase difference between the states of the unit cells, a specular reflection can still be observed. This clearly indicates the influence of other layers within the RIS, beyond the front layer responsible for generating the desired reflection in a specific direction. This suggests that the back layer may be contributing to undesired signal behavior, impacting the overall performance of the system.

\section{Phase Optimization}
\label{sec: phase optimization}

In this section, we focus on optimizing the phase distribution of RIS to maximize the reflection gain at a specifically designed angle. The factors influencing the phase distribution include geometric information like incident and reflection angles, which depend on the transceivers and RIS locations, and RIS hardware parameters such as unit cell size, surface dimensions, spacing between unit cells, phase shift resolution, and other hardware-related effects.  

\subsection{Phase Distribution Error}
\label{sec: phase error}

The phase distribution across the surface significantly impacts the amplitude of the reflected wave, denoted as $g_\mathsf{pd}$. The relationship between the phase distribution and the reflection amplitude is given by\cite{physic_based_ris}:
\begin{equation}
    g_{\mathsf{pd}} = \left|\sum_{n_x=0}^{Q_x-1}\sum_{n_y=0}^{Q_y-1} \mathrm{e}^{\mathrm{j}(\alpha_{n_x,n_y}+\beta_{n_x,n_y})}\right|,
\end{equation}
where $Q_x$ and $Q_y$ are the number of unit cells along $x-$ and $y-$axes, respectively, and $\alpha_{n_x,n_y}$ represents the initial phase of each unit cell, which is determined by the incident and reflection angles, and $\beta_{n_x,n_y}$ is the phase shift applied by each unit cell to achieve the desired reflection.
The initial phase $\alpha_{n_x,n_y}$ for each unit cell is calculated as:
\begin{equation}
    \alpha_{n_x,n_y} = \kappa d_x W_x n_x + \kappa d_y W_y n_y .
\end{equation}
Here, $\kappa = 2\pi/\lambda$ is the wavenumber, $d_x$ and $d_y$ are the distances between unit cells along the $x-$ and $y-$axes, and $n_x$ and $n_y$ are the indices of the unit cells along these axes. The terms $W_x$ and $W_y$ account for the effect of the incident and reflection angles on the initial phase distribution, defined as:
\begin{equation}
    \begin{split}
        W_x & = \sin{(\theta_t)}\cos{(\phi_t)} + \sin{(\theta_r)}\cos{(\phi_r)},\\
        W_y & = \sin{(\theta_t)}\sin{(\phi_t)} +
        \sin{(\theta_r)}\sin{(\phi_r)},
    \end{split}
\end{equation}
where the angle pairs $(\theta_t, \phi_t)$ and $(\theta_r, \phi_r)$ represent the elevation and azimuth angles of the incident and reflected waves, respectively.
To achieve maximum reflection in the desired direction, i.e., $(\theta_t^{\star}, \phi_{t}^{\star})$\footnote{The symbol $\star$ does not denote complex conjugation; it is used solely to indicate a specific reference or target value.} and $(\theta_r^\star, \phi_{r}^{\star})$, each unit cell should apply a phase shift $\beta_{n_x,n_y}$ that precisely cancels out the initial phase $\alpha_{n_x,n_y}^\star$. This phase shift is expressed as:
\begin{equation}
    \beta_{n_x,n_y} = -\alpha_{n_x,n_y}^{\star} = -\kappa d_x W_x^\star n_x - \kappa d_y W_y^\star n_y .
\end{equation}
Following the method outlined in \cite{balanis01} for analyzing uniform planar arrays, and employing mathematical identities for summation and trigonometric functions:
\begin{equation}
    \sum^{N-1}_{n=0}a^n = \frac{1-a^N}{1-a}, \hspace{5mm}
    \sin{(b)} = \frac{\mathrm{e}^{\mathrm{j}b} - \mathrm{e}^{-\mathrm{j}b}}{2\mathrm{j}},
\end{equation}
and setting $a = \mathrm{e}^{\mathrm{j}A}$, we derive:
\begin{equation}
    \begin{split}
        \sum^{N-1}_{n=0}\mathrm{e}^{\mathrm{j}An} &= \frac{1-\mathrm{e}^{\mathrm{j}NA}}{1-\mathrm{e}^{\mathrm{j}A}}
        = \frac{\mathrm{e}^{\mathrm{j}NA/2}}{\mathrm{e}^{\mathrm{j}A/2}}\times\frac{\mathrm{e}^{-jNA/2}-\mathrm{e}^{\mathrm{j}NA/2}}{\mathrm{e}^{-jA/2}-\mathrm{e}^{\mathrm{j}A/2}}\\ &=
        \mathrm{e}^{\mathrm{j}(N-1)A/2}\times\frac{\sin{(NA/2)}}{\sin{(A/2)}}.
    \end{split}
\end{equation}
Applying this identity to compute $g_\mathsf{pd}$, we obtain:
\begin{equation}
    \begin{split}
        g_\mathsf{pd} & = \left| \sum_{n_x=0}^{Q_x-1}\sum_{n_y=0}^{Q_y-1}\mathrm{e}^{\mathrm{j}(\alpha_{n_x,n_y} - \alpha_{n_x,n_y}^\star)} \right|\\ & = \frac{\sin{(Q_x(W_x-W_x^\star)/2))}}{\sin{((W_x-W_x^\star)/2)}} \times \frac{\sin{(Q_y(W_y-W_y^\star)/2))}}{\sin{((W_y-W_y^\star)/2)}}.
    \end{split}
\end{equation}
Any deviation from continuous phase distribution $-\alpha_{n_x,n_y}^\star$ results in a reduction of reflection amplitude in the desired direction and an increase in the phase distribution error. The error of phase distribution $e_\mathsf{pd}$ is quantified as:
\begin{equation}
    \label{epd}
    e_\mathsf{pd} = Q_xQ_y - \left| \sum_{n_x=0}^{Q_x-1}\sum_{n_y=0}^{Q_y-1} \mathrm{e}^{\mathrm{j}(\alpha_{n_x,n_y}^\star + \beta_{n_x,n_y})} \right| .
\end{equation}

\begin{figure*}[t]
    \centering
    \begin{tikzpicture}
        \begin{axis}[
            xlabel={Reflection angle $\theta_r$ [deg]},
            ylabel={Amplitude [dB]},
            legend style={
                at={(1.01,1)}, 
                anchor=north west,
                font=\small, 
            },
            xmin=0, xmax=90,
            ymin=10, ymax=40,
            grid=major,
            width=14cm, 
            height=9cm, 
            thick
            ]

        \addplot[black, thick] table [x index=0, y index=1, col sep=comma] {Data/1bit_3bit_2.csv};
        \addlegendentry{Cont.}
        
        \addplot[red, thick] table [x index=0, y index=2, col sep=comma] {Data/1bit_3bit_2.csv};
        \addlegendentry{3-bit}

        \addplot[green, thick] table [x index=0, y index=3, col sep=comma] {Data/1bit_3bit_2.csv};
        \addlegendentry{2-bit}

        \addplot[blue, thick] table [x index=0, y index=4, col sep=comma] {Data/1bit_3bit_2.csv};
        \addlegendentry{1-bit}

        \end{axis}
    \end{tikzpicture}
    \caption{Amplitude of the response function, $|g_\mathsf{pd}/\lambda|^2$, in dB vs. $\theta_r$ for $(\theta_t^\star,\phi_t^\star) = (0,0)$, $(\theta_r^\star,\phi_r^\star) = (60\text{°},0)$. Reflection comparison of continuous and discrete phase shifter. Low-resolution phase shifter introduces deviations due to the inherent limitations of the discrete phase states.}
    \label{fig: con.vs1-bit}
\end{figure*}
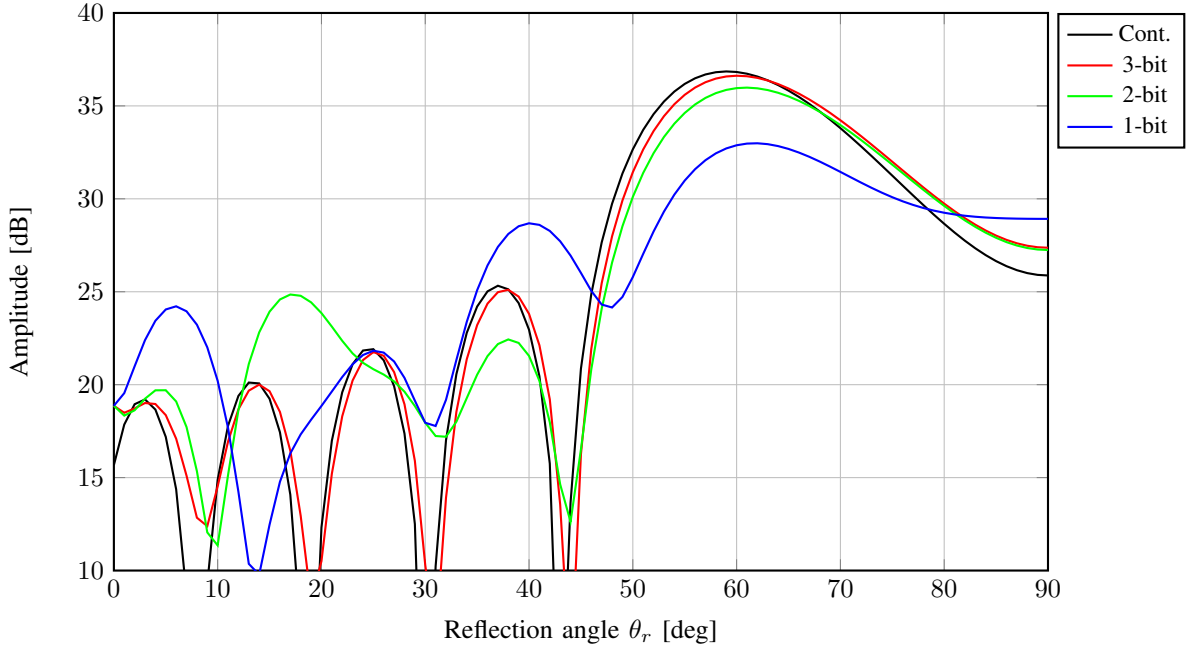

\subsection{Phase Quantization Error}
In this part, we analyze the impact of phase quantization on the reflected signal from the RIS. To illustrate the error, we consider the same exemplary scenario of Section \ref{sec: Natural ref}, and plot the equivalent radiation pattern of the reflecting side vs. the reflection angle in Fig.~\ref{fig: con.vs1-bit}. For comparison, we first consider an ideal continuous phase-shift design, where each unit cell can achieve an arbitrary phase shift without constraints. The continuous-phase RIS exhibits the highest reflection gain at the designed angle. We then examine the performance of quantized phase shifters with varying resolutions.
As depicted in Fig.~\ref{fig: con.vs1-bit}, a 3-bit uniform quantization of the phase shifts produces a response function that closely approximates the ideal continuous-phase case. However, as the quantization resolution decreases, the response function progressively deviates from the ideal case, leading to performance degradation. This observation highlights the trade-off between hardware complexity and reflection efficiency in practical RIS implementations. 

As an example of the quantization process, the following equation illustrates the method used to perform 1-bit quantization based on the continuous phase shifts:

\begin{equation}
    \beta_{n_x,n_y(1\text{-bit})} = 
    \begin{cases}
        \gamma + \pi &  T_1 \leq \beta_{n_x,n_y} \leq T_2 \\
        \gamma & \text{otherwise}
    \end{cases}
    ,
\end{equation}
where $\gamma$ is an arbitrary number, $T_1$ and $T_2$ are quantization thresholds in range of $[0^\circ, 360^\circ)$. As seen in Fig.~\ref{fig: con.vs1-bit}, using the ideal 1-bit phase shifter instead of the continuous phase shifter introduces deviations due to the inherent limitations of the discrete phase states. The 1-bit phase shifter, restricted to ON and OFF states with a 180° phase difference, cannot achieve the same precise control over the phase distribution as the continuous phase shifter. This limitation results in phase quantization errors, which contribute to deviations in the reflected wave's direction and overall performance of the metasurface.

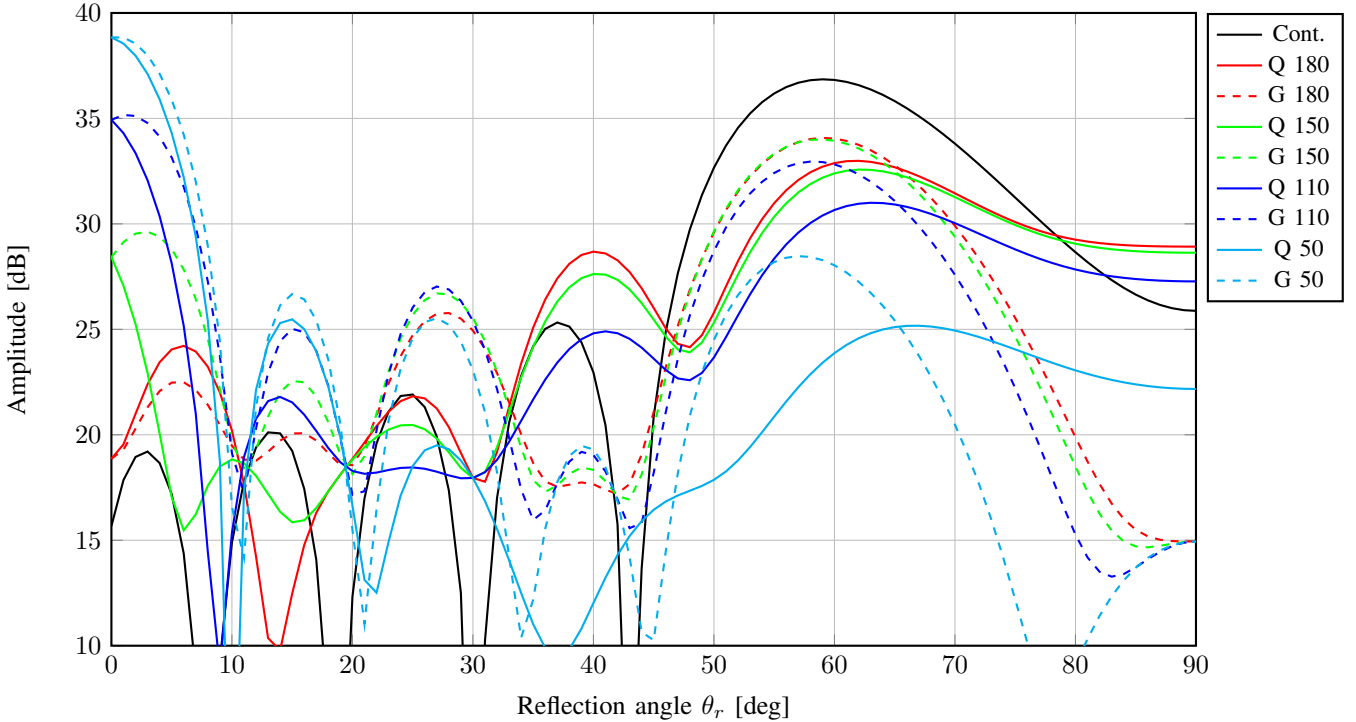
\begin{figure*}[t]
    \centering
    \begin{tikzpicture}
        \begin{axis}[
            xlabel={Reflection angle $\theta_r$ [deg]},
            ylabel={Amplitude [dB]},
            legend style={
                at={(1.01,1)}, 
                anchor=north west,
                font=\small, 
            },
            xmin=0, xmax=90,
            ymin=10, ymax=40,
            grid=major,
            width=16cm, 
            height=10cm, 
            thick
            ]

        \addplot[black, thick] table [x index=0, y index=1, col sep=comma] {Data/GA_QA_compare.csv};
        \addlegendentry{Cont.}
        
        \addplot[red, thick] table [x index=0, y index=2, col sep=comma] {Data/GA_QA_compare.csv};
        \addlegendentry{Q 180}

        \addplot[red, thick, dashed] table [x index=0, y index=3, col sep=comma] {Data/GA_QA_compare.csv};
        \addlegendentry{G 180}

        \addplot[green, thick] table [x index=0, y index=4, col sep=comma] {Data/GA_QA_compare.csv};
        \addlegendentry{Q 150}

        \addplot[green, thick, dashed] table [x index=0, y index=5, col sep=comma] {Data/GA_QA_compare.csv};
        \addlegendentry{G 150}

        \addplot[blue, thick] table [x index=0, y index=6, col sep=comma] {Data/GA_QA_compare.csv};
        \addlegendentry{Q 110}

        \addplot[blue, thick, dashed] table [x index=0, y index=7, col sep=comma] {Data/GA_QA_compare.csv};
        \addlegendentry{G 110}

        \addplot[cyan, thick] table [x index=0, y index=10, col sep=comma] {Data/GA_QA_compare.csv};
        \addlegendentry{Q 50}

        \addplot[cyan, thick, dashed] table [x index=0, y index=11, col sep=comma] {Data/GA_QA_compare.csv};
        \addlegendentry{G 50}

        \end{axis}
    \end{tikzpicture}
        \caption{Comparison of the reflection patterns for continuous, 1-bit quantized, and optimized phase shifters by GA, demonstrating the reduction in gain degradation at the desired angle (i.e., 60°) achieved through the optimization approach. (In legend table, Cont. = Continouos phase shift, Q $\psi$ = Quantized with $\psi$° phase difference, and G $\psi$ = Optimized by GA with $\psi$° phase difference.)}
    \label{GA comparison}
\end{figure*}

\subsection{Genetic Algorithm (GA)}
 GA is a heuristic search method, a robust evolutionary technique suitable for discrete, non-convex optimization tasks inspired by natural selection, making it particularly effective for solving complex, high-dimensional optimization problems where traditional methods may struggle. It operates by generating an initial population of candidate solutions, each encoded as a chromosome. These solutions are iteratively refined through selection, crossover, and mutation, with their fitness evaluated based on a predefined objective function \cite{Mitchell1996AnIT}.

GA offers a good trade-off between computational efficiency and solution quality. Its population-based nature allows for better exploration of the binary search space, and the flexibility to incorporate hardware constraints directly into the gene space makes it well-suited for RIS design. Furthermore, adaptive mutation and elitism strategies improve convergence stability. For these reasons, GA was selected as the core optimization method in this work.

To mitigate the deviations caused by the use of a 1-bit phase shifter, we optimized the phase distribution across the surface using a GA. The objective is to find a binary phase distribution that maximizes the coherent summation of reflected signals across the surface. 

The fitness function quantifies the effectiveness of each candidate solution in optimizing the phase distribution of the RIS. Each solution modifies the initial phase matrix, and its quality is determined by the constructive or destructive interference it introduces.
Let us denote $\alpha^{\star}_{n_x,n_y}$ the initial phase of unit cells and $\Phi_{n_x,n_y}$ represent the binary value assigned to each unit cell by the optimization algorithm. The objective is to maximize the magnitude of the sum of elements which leads to minimizing the error of phase distribution $e_\mathsf{pd}$ (\ref{epd}):
\begin{equation}
    \max_{\Phi\in\{0,1\}^{Q_x\times Q_y}} \left| \sum_{n_x=0}^{Q_x-1}\sum_{n_y=0}^{Q_y-1} \mathrm{e}^{\mathrm{j}(\alpha^{\star}_{n_x,n_y} + \psi\Phi_{n_x,n_y})} \right|
\end{equation}
where $\psi$ is the phase difference between the ON and OFF states of the unit cells. The fitness function evaluates the magnitude of the complex sum resulting from the superposition of the hardware-induced and optimized phases. To balance exploration and exploitation, the algorithm integrates tournament selection, single-point crossover, and adaptive mutation strategies, while retaining top solutions through elitism. The population size is selected based on the number of unit cells in the RIS and the overall complexity of the optimization problem, with early stopping triggered when no significant improvement is observed.

\par To assess the performance of the optimized phase distribution, we compared three configurations: (i) a continuous phase shifter (serving as the ideal upper bound), (ii) a 1-bit quantized phase shifter, and (iii) a 1-bit optimized phase shifter using GA. The RIS was illuminated with a normally incident wave, with the objective of achieving a reflection at 60°. The GA was specifically designed to minimize gain degradation at the desired reflection angle.

As illustrated in Fig.~\ref{GA comparison}, the continuous phase distribution represents the upper bound of achievable reflection gain at the designed angle, with minimal specular reflection due to the absence of phase shift constraints. In contrast, a 1-bit quantized phase shifter with a 180° phase difference between ON and OFF states reduces the gain by 4 dB at 60°. Furthermore, in the quantized approach, the peak of the main lobe is no longer exactly aligned with the designed angle. However, optimizing the phase distribution while maintaining this phase difference recovers 2 dB of the lost gain and realigns the peak of the main lobe with the designed angle.
                                          
Optimization techniques, such as GA, improve reflection performance at the designed angle, but do not mitigate specular reflection, as the latter is fundamentally constrained by the phase shift limitation rather than the phase distribution. Moreover, practical hardware limitations often prevent achieving the ideal 180° phase difference. To address this, we examine non-ideal cases and adjust the phase distribution accordingly.
For a 150° phase difference (green curve), the quantized case suffers a performance loss compared to 180°. However, optimization allows the system to closely approach the optimized 180° performance, albeit at the cost of increased specular reflection at 0°, stemming from hardware constraints. With a 110° phase difference (blue curve), the degradation in the quantized case is more pronounced, but optimization restores performance to a level comparable to the 180° quantized scenario. This demonstrates that a phase deviation can be mitigated through proper phase distribution, highlighting the critical role of optimization.
Finally, for a 50° phase difference, the results further validate the effectiveness of optimization while reinforcing a key observation: although optimizing the phase distribution enhances reflection performance, it does not reduce specular reflection. This limitation arises because both quantized and optimized distributions share the same fundamental phase differences between unit cell states, resulting in comparable specular reflection levels.

Nevertheless, although our optimization process primarily aims to enhance reflection at the designed angle, it also reduces the closest side lobe to the main lobe. This indicates that the optimization not only improves the main reflection but also provides greater control over the side lobes, thereby increasing the flexibility of the overall reflection response of the RIS. Further investigation into phase distribution optimization for simultaneous control of side lobes and the main lobe remains an open research question.

\section{Measurement Setup and Results}
\label{sec:Measurement}

\begin{figure}[t]
    \centering
    \includegraphics[width=1\linewidth]{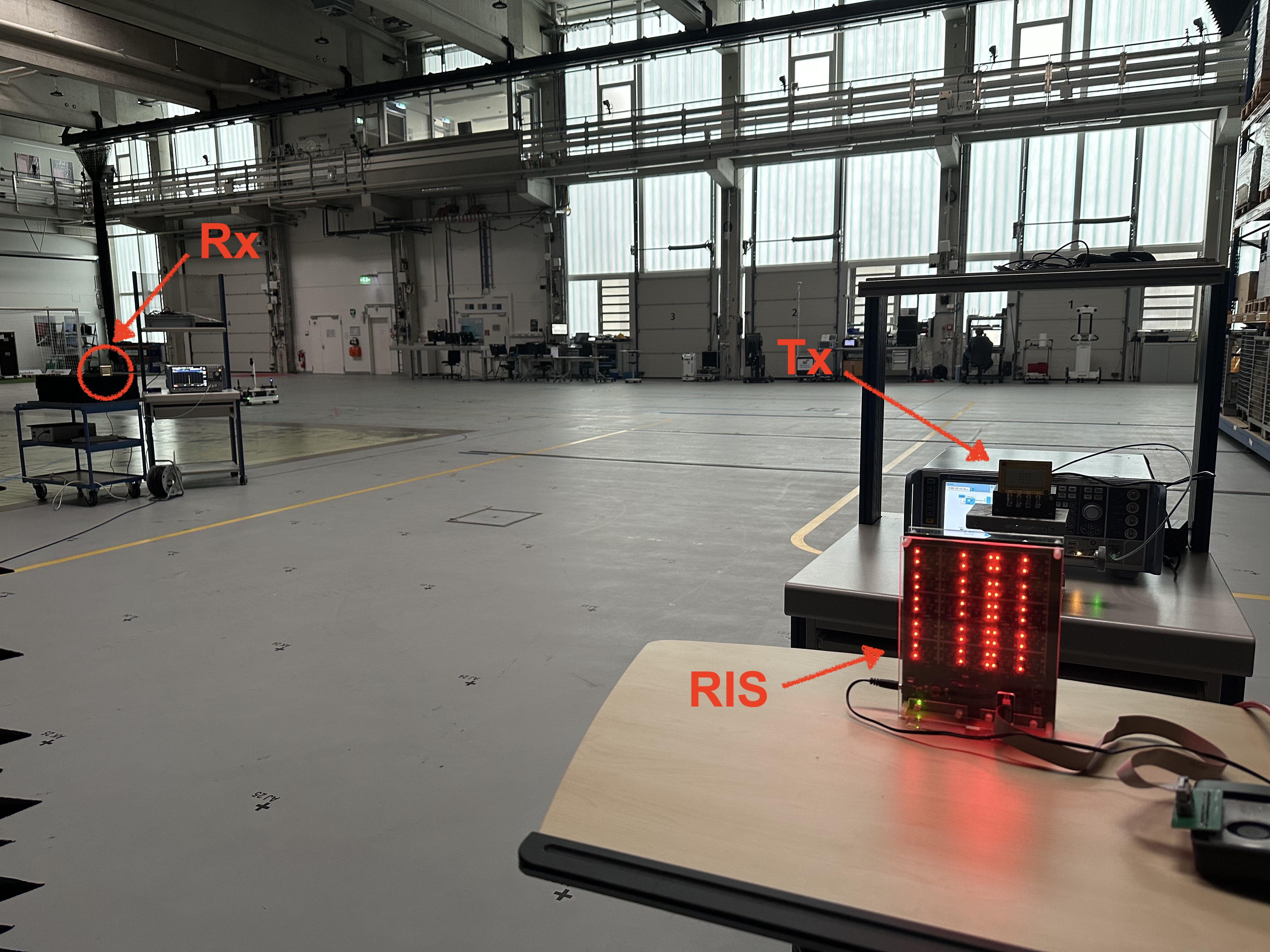}
    \caption{Testbed setup at the LINK Test Center demonstrating the configuration of the transmitter, receiver and RIS operating at 28 GHz, with the incoming wave impinging normal (0°) on the RIS and its controlled reflection at 60°.}
    \label{measurement setup}
\end{figure}

To validate our hardware implementation and optimization approach, we conducted measurements using the setup illustrated in Fig.~\ref{measurement setup}. The experiment was performed at the LINK Test Center in Nuremberg \cite{Fraunhofer-IIS}.

In our measurement configuration, the incident wave impinges normal to the RIS's surface (0°), while the phase distribution of the RIS is designed to reflect the wave at an angle of 60°. The receiver (Rx) is placed in the far-field region of the RIS and at the designated reflection angle of 60°, aligning with the intended phase design.

The RIS used in this experiment consists of 100 unit cells arranged in a $10 \times 10$ grid, with each element having a size of $\lambda/2$ in both dimensions. As a 1-bit RIS, each unit cell can switch between two distinct states. To enhance visualization, each unit cell is equipped with an indicator light: when the unit cell is active (ON), the corresponding light illuminates in red. This feature allows for an intuitive real-time representation of the RIS phase configuration.

The primary objective of this measurement is to analyze the impact of different RIS phase distributions on wave reflection and received power, while keeping all other components of the measurement setup constant. To this end, we evaluate two distinct phase configurations: (i) a phase distribution obtained through a quantization-based method and (ii) an optimized phase distribution derived using a GA. By comparing these approaches, we aim to assess the effectiveness of optimization techniques in enhancing RIS performance.

\begin{figure}[t]
    \centering
    \begin{subfigure}[b]{0.25\textwidth}
        \centering
        \includegraphics[width=\textwidth]{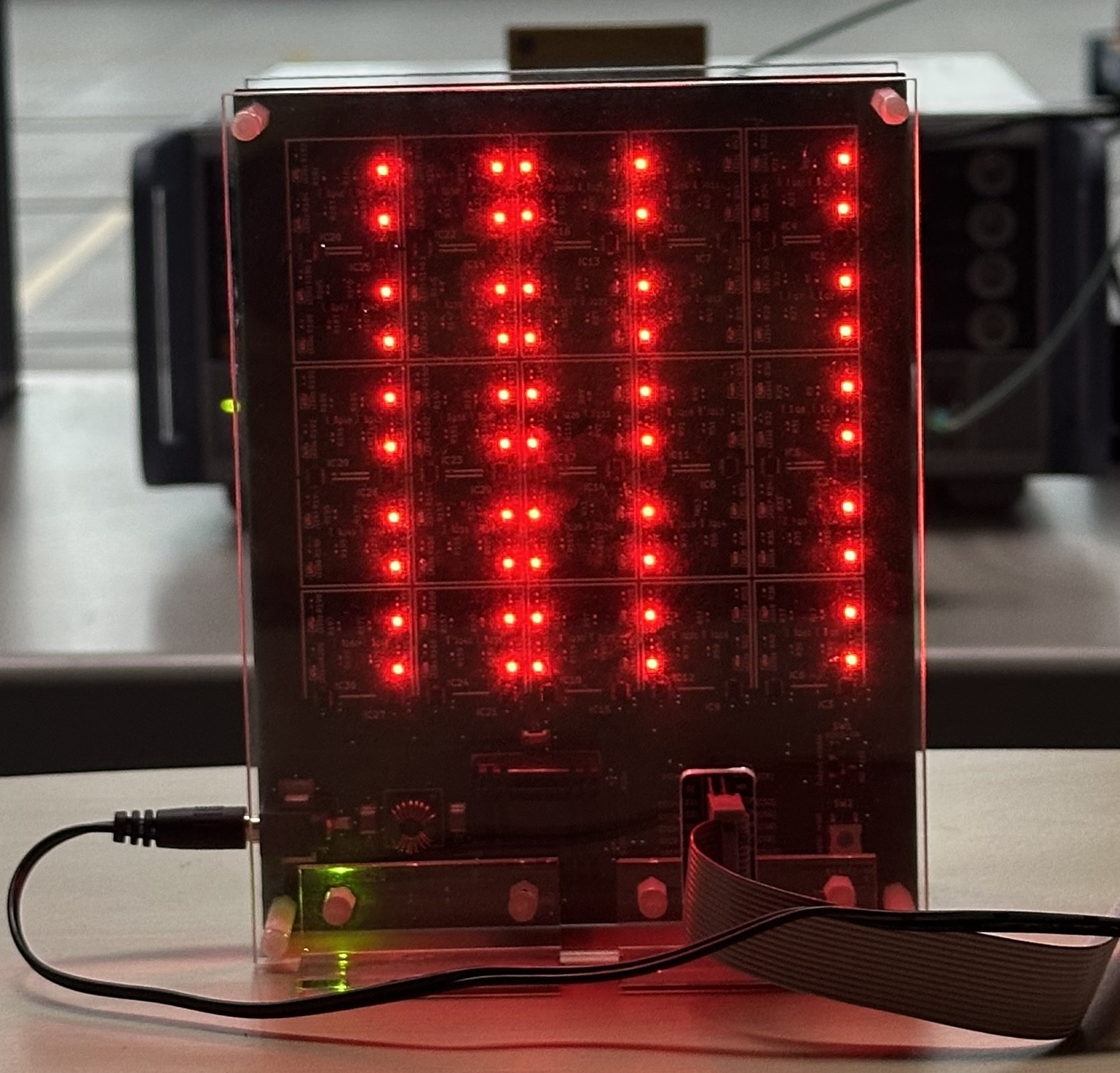}
        \caption{}
        \label{fig:sub1}
    \end{subfigure}
    \hfill
    \begin{subfigure}[b]{0.25\textwidth}
        \centering
        \includegraphics[width=\textwidth]{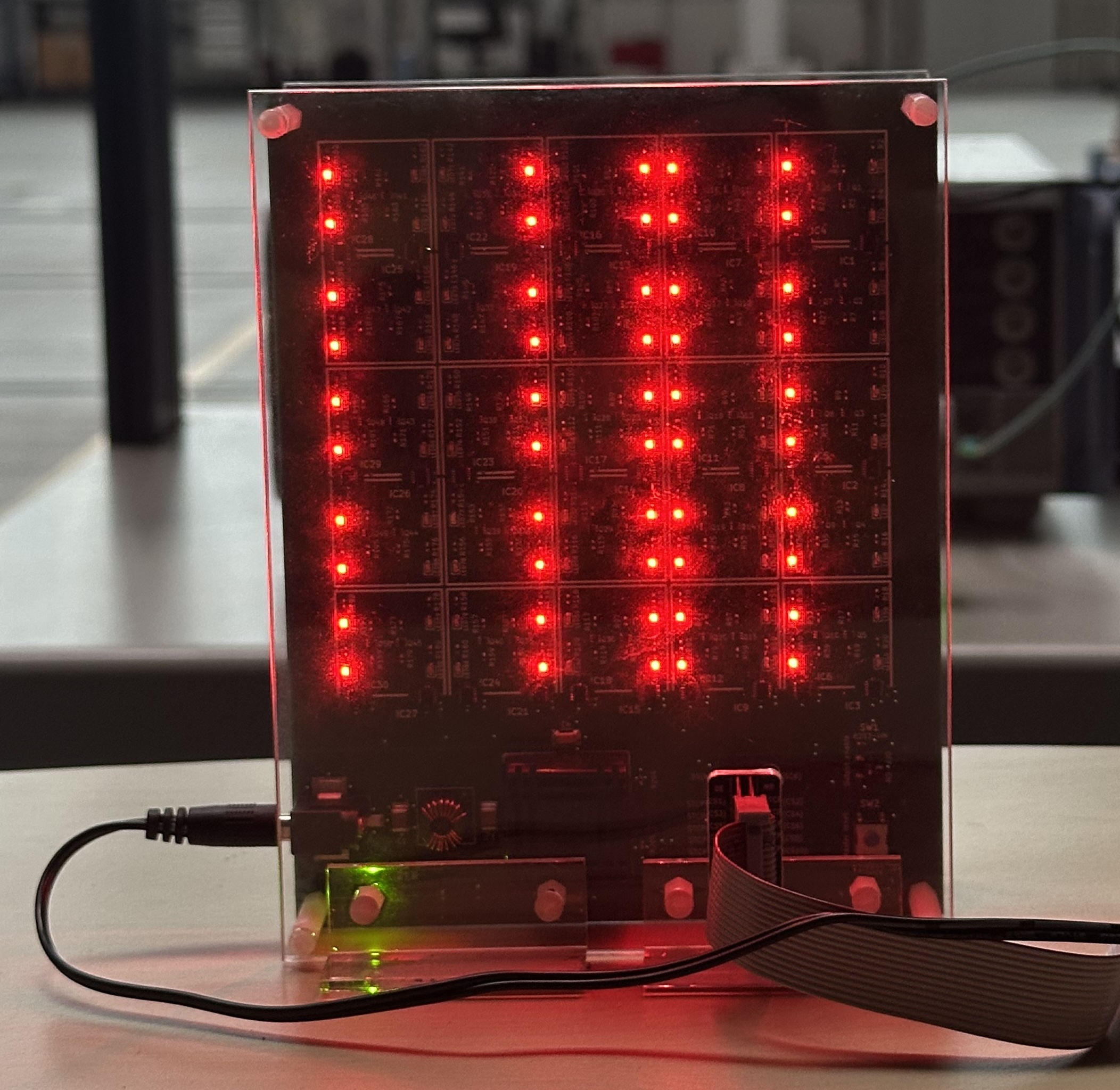}
        \caption{}
        \label{}
    \end{subfigure}
    \caption{Phase distributions obtained using (a) the quantization method and (b) the GA-optimized method, yielding a 2.82 dB increase in received power with the optimized distribution.}
    \label{fig:quantized}
\end{figure}

Fig.~\ref{fig:quantized} presents the phase distributions obtained using the quantization method and the GA-optimized method. The phase distributions exhibit a distinct difference, which highlights how the GA adapts the phase shifts to better align with the desired reflection pattern.
The most significant finding of the measurement is the improvement in the strength of the received signal. The optimized phase distribution obtained through GA results in a 2.82 dB increase in received power compared to the quantized approach, demonstrating the effectiveness of optimization in overcoming the limitations of discrete phase shifts.
The reason for this improvement lies in the way GA fine-tunes the phase distribution. In the quantization-based method, phase errors accumulate across the RIS, leading to destructive interference and energy leakage in unintended directions. In contrast, GA carefully selects phase values to constructively align reflected waves at the receiver’s location, maximizing signal strength. This confirms that even with a 1-bit RIS, optimization can significantly improve system performance.
These results emphasize the importance of advanced optimization techniques in RIS design. While simple quantization provides a feasible baseline, sophisticated algorithms such as GA unlock greater potential, making RIS more effective for practical applications.

\section{Conclusion}
\label{sec: conclusion}

This work presented the development and experimental evaluation of a 1-bit RIS prototype operating at 28 GHz, addressing the impact of hardware limitations on RIS performance. Unlike conventional studies that model RIS as an ideal phase-shift matrix, this study considered real-world constraints, including specular reflection caused by hardware limitations and phase quantization errors. A GA-based optimization was proposed to mitigate quantization-induced performance degradation, enhancing the reflection pattern and maximizing reflection power.
Experimental measurements validated the effectiveness of the proposed approach, demonstrating significant improvements in RIS performance despite hardware constraints. The findings emphasize the importance of advanced phase optimization techniques in practical RIS implementations, particularly in mmWave scenarios where hardware limitations play a crucial role.
As RIS technology continues to evolve, incorporating advanced optimization techniques will be crucial in overcoming current hardware constraints and achieving higher levels of control over EM wave reflections.
Future research could explore extending the optimization framework and further investigating the optimization methods to enhance performance. Additionally, experimental validation in dynamic environments could provide deeper insights into real-world deployment challenges.

\section*{Acknowledgment}
This work has been funded by the Federal Office for Information Security (BSI) within the project RIS4NGWB under the grant identifier 01MO23001A-B.

\bibliographystyle{IEEEtran}
\bibliography{IEEEabrv,bibfile}
\end{document}